\documentclass[conf]{new-aiaa}
\usepackage[utf8]{inputenc}
\usepackage{graphicx} 
\usepackage{xcolor}
\usepackage{graphicx}
\usepackage{amsmath}
\usepackage[version=4]{mhchem}
\usepackage{siunitx}
\usepackage{longtable,tabularx}
\usepackage{textcomp}
\newcommand{\colorann}[3]{\textcolor{#1}{${}^{#2}[$#3$]$}}
\newcommand{\robyn}[1]{\colorann{red}{robyn}{#1}}
\newcommand{\myra}[1]{\colorann{red}{myra}{#1}}

\title{A Family-Based Approach to Safety Cases for Controlled Airspaces  in Small Uncrewed Aerial Systems}

\author{Michael C. Hunter,\footnote{Graduate Research Assistant, 115 Atanasoff Hall, Iowa State University, Ames, IA, USA}  Usman Gohar,\footnote{Graduate Research Assistant, 116 Atanasoff Hall, Iowa State University, Ames, IA,  USA} Myra B. Cohen,\footnote{Professor, Computer Science, 202 Atanasoff Hall, Iowa State University, Ames, IA,  USA} Robyn. R. Lutz \footnote{Professor, Computer Science, 230 Atanasoff Hall, Iowa State University, Ames, IA, USA}}
\affil{Iowa State University, Ames, IA, USA}

\author{Jane Cleland-Huang\footnote{Professor, Computer Science and Engineering, 325B Fitzpatrick Hall, University of Notre Dame, IN USA}}
\affil{Notre Dame University, Notre Dame, IN, USA}

\begin{document}

\maketitle

\begin{abstract}
As small Uncrewed Aircraft Systems (sUAS) increasingly operate in the national airspace, safety concerns arise due to a corresponding rise in reported airspace violations and incidents, highlighting the need for a safe mechanism for sUAS entry control to manage the potential overload. This paper presents work toward our aim of establishing automated, customized safety-claim support for managing on-entry requests from sUAS to enter controlled airspace. We describe our approach, Safety Case Software Product Line Engineering (SafeSPLE), which is a novel method to extend product-family techniques to on-entry safety cases. It begins with a hazard analysis and design of a safety case feature model defining key points in variation, followed by the creation of a parameterized safety case. We use these together to automate the generation of instances for specific sUAS. Finally we use a case study to demonstrate that the SafeSPLE method can be used to facilitate creation of safety cases for specific flights.  
\end{abstract}

\section {Introduction}
In recent years, we have seen a rise in the use of small Uncrewed Aircraft Systems (sUAS), which are increasingly deployed into shared areas of the national airspace. This growth is being driven by commercial, recreational, and service applications such as package delivery, photography, remote sensing, and emergency response \cite{Erdelj2017HelpFT,FAA}. Unfortunately, there has also been an increase in reported incidents involving sUAS, often linked to hardware or software malfunctions and human errors, thus increasing the risk of serious accidents \cite{cleland2022towards}. Additionally, external elements such as radio interference and adverse weather conditions have been identified as contributing factors in many reported incidents \cite{FAA,report3}.  This is leading to a need for a clearer understanding and regulations about when sUAS can participate in shared airspace. Ultimately, for controlled airspaces, both the sUAS and its operator will need vetting prior to entry to ensure the safety of all within close proximity \cite{gohar2024towards}.

Safety cases to certify safety-critical systems are a common practice worldwide, including for space, defense, rail, nuclear, healthcare, and oil and gas systems \cite{NASA12,Knig12, Hatcliff14, Leveson23}. 
In commercial aviation, certification of a plane takes a long period of evaluation. As we transition to the more dynamic scenario of many different types of sUAS flying in shared airspace, with tens or hundreds of thousands of variations in the way each sUAS, its operator, its mission, and its operating environment can be combined \cite{10.1145/3544548.3581003}, the idea of building bespoke safety cases becomes futile. Organizations or companies deploying sUAS are less likely to have the staff or expertise to develop these documents. Additionally, many sUAS are operated by individuals (perhaps a hobbyist) wanting to perform an ad hoc flight, for example, to take photos. 

One way to begin to address these concerns is an approach being taken by the U.S. Federal Aviation Administration (FAA) and the National Aeronautics and Space Administration (NASA). They are collaborating to develop a UAS Traffic Management (UTM) system where sUAS can be dynamically certified to enter (or denied entry to) the shared airspace \cite{UTM}.  

If we can design a set of criteria that can be quickly evaluated, along with requirements for 
real-time (temporary) certification, it will allow automated decision-making for entry into a UTM.  While this vision of a dynamically controlled UTM is still in development, we argue there are approaches that may help to streamline such a system and can lead to re-use and efficiency in the eventual UTM.

In this work, we propose the use of families (software product lines) of safety cases, which represent both the common and the variable features of flight characteristics (such as the operator, weather conditions, vehicle types, etc.), that will cover all possible instances of a safety case for these features. We note that these features are unlike traditional features, which normally represent program functionality. Instead, they can be mapped to the context of a safety case, which restricts the conditions under which the safety case arguments hold.  We propose that these possible alternatives can be mapped to key points of variation in a parameterized, general safety case for controlled airspace. Further, we propose that a customized instance can be automatically generated from the product line in conjunction with a parameterized safety case for each sUAS seeking entry to that airspace. 
As part of the safety case, the operator interface will detail what evidence is needed to satisfy the safety goals, and, assuming the sUAS can produce this evidence, it will be allowed to enter. Without that evidence, entry will be denied.
Referred to as SafeSPLE (or Safety Case Software Product Line Engineering), the core of our approach lies in the use of a Safety Case Software Product Line (SafeSPL).

\section{SafeSPLE}
\label{sec:SafeSPLE}

\begin{figure}[ht]
\centering
\includegraphics[width=6.3in]{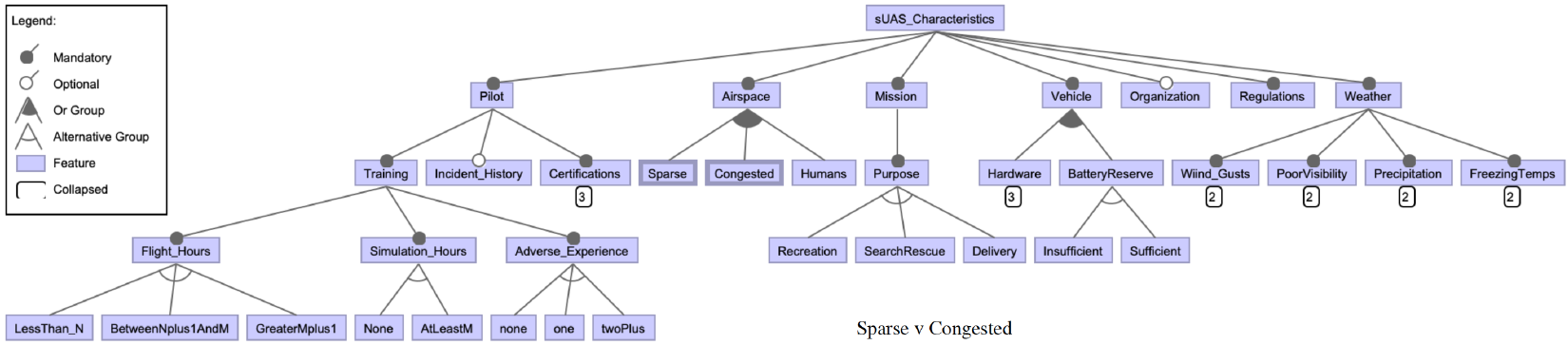}
\caption{Partial feature model representing heuristics for a safety-case software product line with 51 features. This represents 288,202 different variants -- combinations of flights, vehicles, operators, and weather conditions. Additional variants such as terrain and its altitude are not  shown.}\label{fig:featuremodel}
\end{figure}

A Software Product Line (SPL) represents a family of software-intensive systems that share a common, managed set of features developed from a common set of core assets in a prescribed way \cite{Clements2001, weiss1999software, Pohl-PL-Eng-Book, Kang1990, SEI-PL}. Individual products are constructed by selecting a set of alternative and optional features (variabilities) and composing them on top of a set of common base features (commonalities) \cite{DonBatory-FeatureDef, pl-feature-modeling-kang, DBLP:books/daglib/0087788, Gomaa, Fantechi-Gnesi}. In the case of a SafeSPL, features representing groupings of mandatory and optional safety-case nodes are combined and configured to generate a valid and appropriate safety case in support of each unique on-entry request. 

The features are typically represented using a feature model, which describes all possible valid combinations or instances. Figure \ref{fig:featuremodel} shows part of an example feature model that we developed for an sUAS mission. It has 51 features and represents 288,202 different variants, 
where each of those variants is a valid possible configuration of an sUAS. In a feature model such as this, the features are shown as rectangles, with a set of dependencies (or constraints) between them. In our example, we have features for the pilot, airspace, vehicle, and weather, among others. For some features, we use ``XOR" alternative conditions (e.g., the purpose of the flight is recreational, search and rescue, or delivery), and in others, we have ``Or" conditions that allow for one or more of the features to be selected (e.g., the airspace can be sparse or congested, and the ground below the airspace may have, or not have, human activity). We also include a \textit{cross-tree constraint} stating that we may have either Sparse or Congested airspace (but not both), which allows either of those airspace descriptions to include human activity. 
Given that the feature model describes all possible valid combinations and cross-tree constraints that can be represented by first-order logic, we can potentially use satisfiability solvers to reason about a product's validity or perform analyses on subsets of the product. There are also existing tools that can convert this representation into logic to help with reasoning~\cite{BENAVIDES2010615,fama}. 
Once the feature model is in a logical representation, we can ask questions about the valid number of products, whether individual products are valid, and/or evaluate the size of a \textit{slice} through the feature model (all products given the selection of a specific concrete feature)~\cite{fama,featureide,10.1145/3176644}. We can also use this to generate test cases for the product line \cite{5456077}.

\begin{figure}[]
\centering
\includegraphics[scale = 0.70]{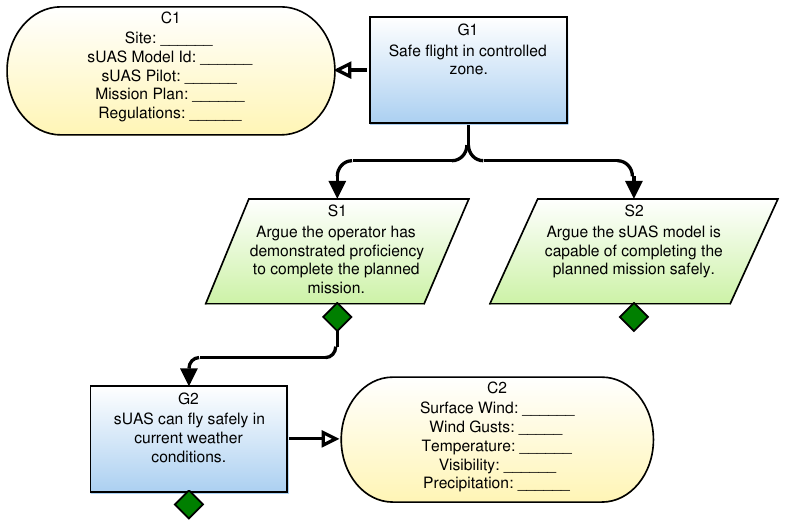}\caption{Partial parameterized safety case. Represents the top level goal (G1) and strategies. The Context nodes (C1 and C2) have parameters which can be instantiated with a defined set of concrete values.}\label{fig:param}
\end{figure}

Figure \ref{fig:param} shows a small fragment of a parameterized safety case. This is based on recent work on assurance recipes \cite{FirestoneC18} and safety patterns \cite{depai2016}, which provide a way to generalize common parts of a safety case. In this figure, we have a primary, top-level Goal (G1) to ensure ``safe flight in a controlled zone.'' It is parameterized by its Context (C1), which typically must be verified for a range of concrete conditions. These variables will also be propagated to the solution (or evidence) nodes. The airspace where this sUAS will fly, the model of sUAS hardware, the operator (or pilot), the planned mission, and any specific regulations placed on this airspace are listed as variables. We next show two high-level strategies (but only expand on one). Strategy S1 links the top-level goal G1 to a subgoal (G2) of flying safely within its current wind conditions. The context for this strategy (C2) has parameters related to the surface wind, gusts, temperature, visibility, and precipitation. At each step, as we move down this argumentation structure, we can utilize the sUAS's specific, concrete parameters to plug in the specific goals/strategies and/or concrete evidence for its customized safety-case instantiation. Part of this process will require the definition of equivalence classes (or choices) which represent ranges of (or discrete) values that behave in a similar way.
 
We recently surveyed stakeholders about the types of features that they consider to be most important when designing a UTM-type on-entry system \cite{gohar2024towards}. Study participants were asked to approve or reject flight-entry requests presented in the form of vignettes. Participants assessed the importance of pilot, drone, environmental, and mission concepts, as well as providing textual feedback. Responses from this initial survey of ours showed that flight characteristics and environmental conditions were seen as most important, and that pilot and drone capabilities should also be considered. Textual feedback relayed doubts about any AI usage, the importance of Human-on-the-Loop, and the need for transparency in automated decision-making. Survey feedback on the set of appropriate criteria to use for on-entry decisions tended to confirm the features in our preliminary sUAS feature model, and the results will inform our design decisions for SafeSPLE going forward.

\section{Vision and Process}

\begin{figure}[ht]
\centering
\includegraphics[width=6.3in]{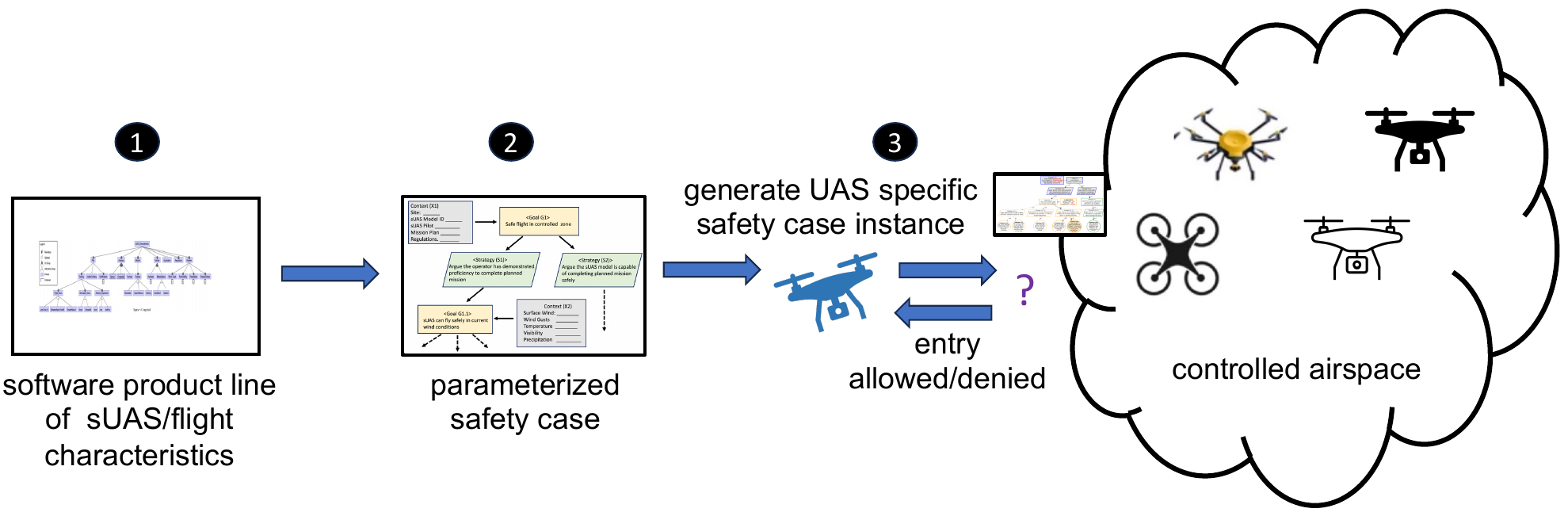}\caption{Overall vision for SafeSPLE. Step one creates a feature model for the safety case product line. Step two designs the parameterized safety case, and Step three generates an instance for a specific sUAS and its characteristics (pilot, vehicle, mission, weather, etc.)}\label{fig:vision}
\end{figure}

Putting this all together, Figure \ref{fig:vision}  presents our vision for SafeSPLE. 
We begin with a software product line based on features such as the sUAS, flight plans, operator and weather conditions, etc.  We input this information to our parameterized safety case, which can be used to generate individualized safety cases for specific sUAS \cite{VierhauserBWXCH21} as they request entry to the UTM. We use the features of our feature model to parameterize the context and evidence. Each instance is a detailed safety case that is valid for that specific instance of the product line under the given context. Since we may not have all evidence required by the generated evidence nodes, a key part of this process will involve determining appropriate (i.e., safe) approximation for reuse across the evidence nodes \cite{AgrawalKVRCL19}.  
Moreover, since each sUAS's safety case will need only a portion of the baseline safety case's options, we are hopeful that performance will be satisfactory for computing in real time.

We now describe the steps taken during the SafeSPLE process as shown in Figure \ref{fig:vision} .

\begin{enumerate}
\item \textbf{Tracing hazards to sUAS features.} We first identify and prioritize the hazards for a SafeSPL using previous studies \cite{FAA} and our surveys \cite{gohar2024towards}. We then associate hazards with the features that have been contributing causes to those hazards' occurrences, as well as associating them with those features' variation points (Figure \ref{fig:featuremodel}). To assure  coverage  of all relevant features in the SafeSPL we also identify any features involved in mitigations. 

\item\textbf{Parameterizing safety-case variability.} 
We next build the parameterized safety case (Figure \ref{fig:param}), modeling it with a combination of feature modeling and GSN \cite{GSN} tools to visualize and reason about valid products and avoid conflicts among features.\footnote{We used the FeatureIDE\cite{featureide} and AdvoCATE\cite{advocate} tools in the this paper.} As part of this step, we create equivalence classes for parameters and map these to the features in the SafeSPL.

\item\textbf{Generating sUAS-specific safety-case instances.} The final step is to generate on-demand instances of the safety case with each element being tagged with a unique, machine-readable link to its position in the baseline safety case for traceability. We also generate a list of required evidence from the leaf nodes to allow the sUAS to provide sufficient safety information.  If entry is denied the traceability will provide partial explanations for the operator. 
\end{enumerate}

\section{Case Study on SafeSPLE}
We now demonstrate via a case study one way to implement a SafeSPL and parameterized safety cases.  The first part of our process is a hazard analysis. We then build a feature model. The features are then used to parameterize our safety case. Lastly, we can generate safety-case instances as requested for any of the concrete combinations of features.  

\subsection{Hazard Analysis}
To begin the SafeSPLE process for a UAS flight, we analyze the hazards of that flight, which is an important first step before creating a safety case \cite{Knig12}. A hazard is a state or event that can potentially result in an accident \cite{ericson2015hazard}. In this work, we do not describe this part of the process in depth but rather list a few of the key hazards we identified. We utilized several sources to create our list of hazards. First, we referenced several papers describing hazard analysis or safety cases for sUAS flights \cite{denpai2016, clodenpai2017, sora}. Next, we discussed sUAS hazards with colleagues and experts who have studied sUAS and flown them. This investigation gave us an extensive list of hazards, which was too long and broad to include in this paper. We narrowed down this extensive list to focus on the following hazards. 

\begin{itemize}
    \item Too much precipitation
    \item Insufficient visibility
    \item Temperatures outside the operating specifications of the sUAS
    \item Wind gusts outside the operating specifications of the sUAS
    \item Insufficient battery for the mission
\end{itemize}

The hazards above are not intended to be fully described or defined, and we do not include prevention or recovery controls or escalation factors for any of these hazards (see \cite{denpai2016} for a more in-depth discussion of hazard analysis). The ultimate consequences of each of the above hazards are generally either loss of separation from the ground or loss of separation from other air traffic. Either of these consequences could lead to the destruction of property, injury, or death. A complete risk analysis of these consequences is likewise beyond the scope of this paper. We illustrate our family-based approach below using a subset of the identified hazards in order to show how the parameterized safety case addresses the hazards for different sUAS.       

\subsection{Feature Model}

\begin{figure}[ht]
    \centering
    \includegraphics[width=.8\textwidth]{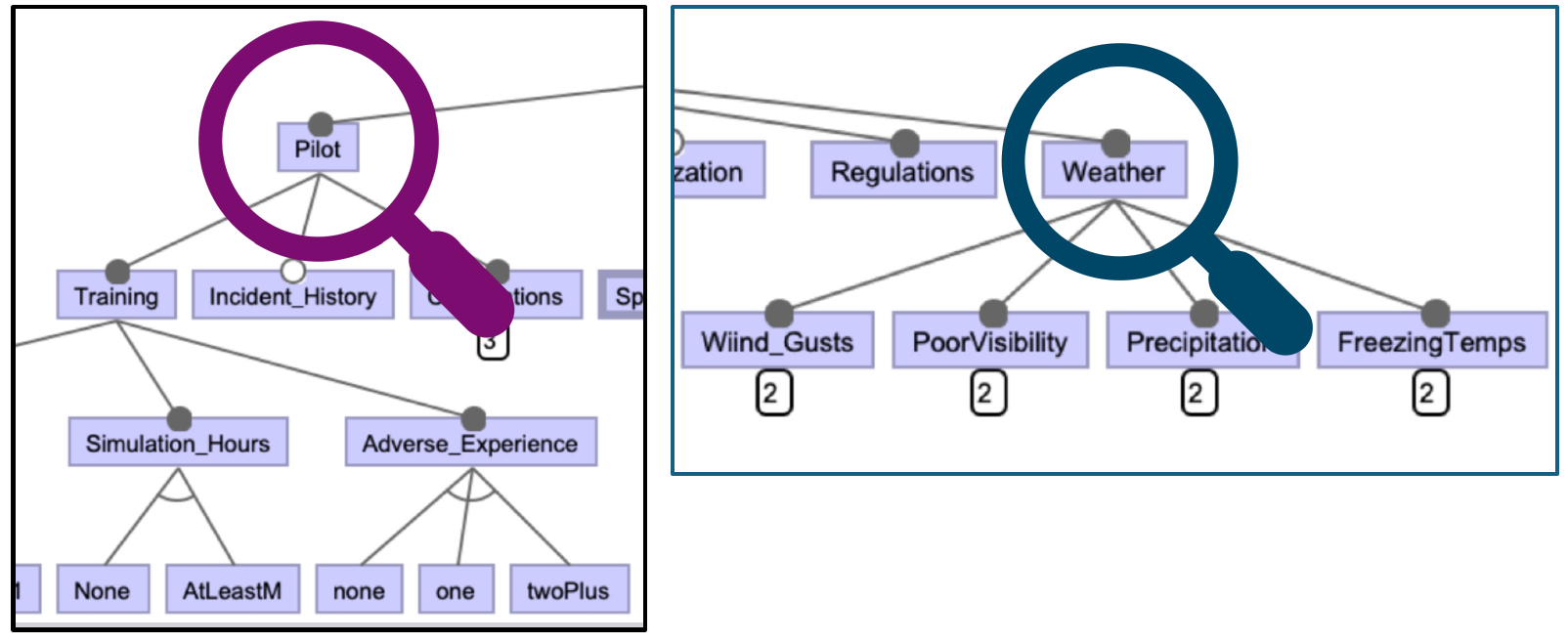}
    \caption{Two parts of the feature model that we focus on for this case study.}
    \label{fig:feature_model_focus}
\end{figure}

The next step in the SafeSPLE process is to create a partial feature model that could apply to a wide variety of sUAS models and missions in controlled airspace as described in Section \ref{sec:SafeSPLE} and Figure \ref{fig:featuremodel}. Since this feature model includes information about the pilot, airspace, mission, vehicle, and weather (among other things), it allows for a wide variety of different types of parameters to be used in our parameterized safety case. In figure \ref{fig:feature_model_focus} we show the two parts of the feature model that are the focus of our safety cases here - the pilot and the weather. These parameterized safety cases are described in the next section.

\subsection{Parameterized Safety Case}

\begin{figure}[ht]
    \centering
    \includegraphics[width=.7\textwidth]{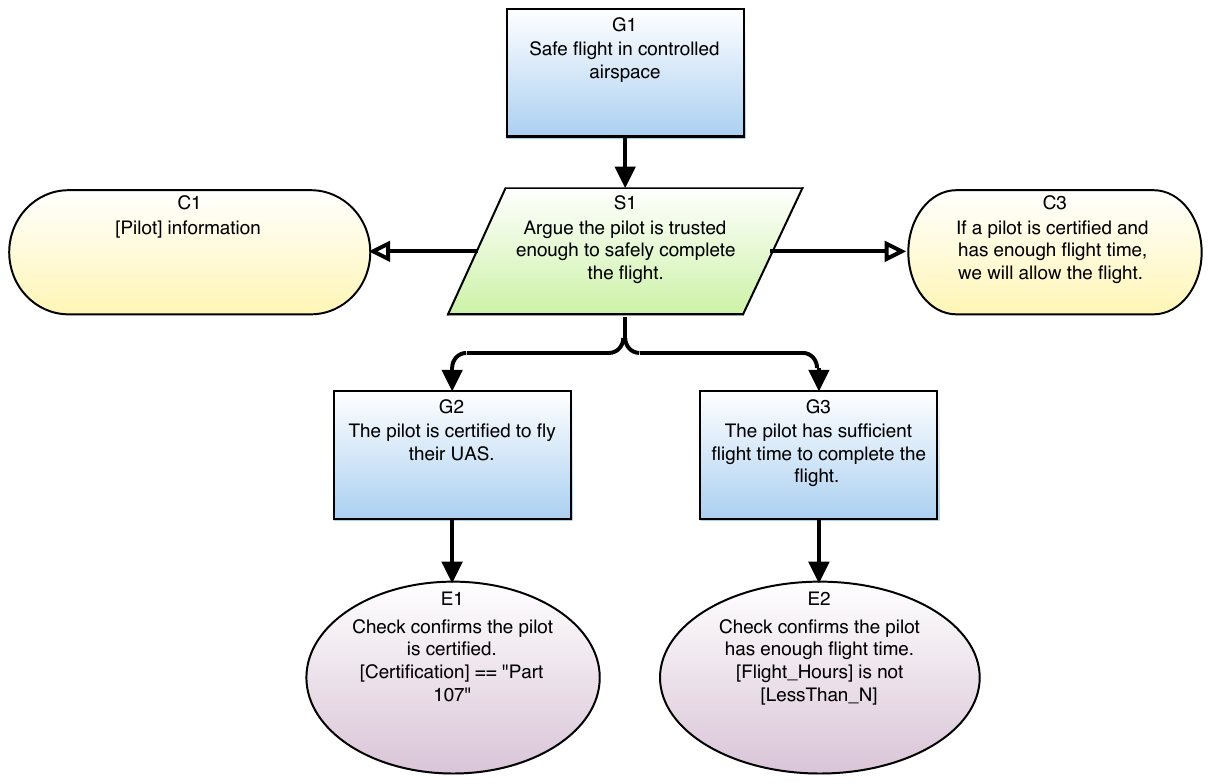}
    \caption{Pilot Safety Case: A safety case based only on whether the pilot is certified and has sufficient experience.}
    \label{fig:pilot_only}
\end{figure}

The next step in our case study (based on SafeSPLE) is to create two illustrative parameterized safety cases for our controlled airspace. The first safety case, seen in Figure \ref{fig:pilot_only}, is based solely on the pilot. It checks whether the pilot is certified and has sufficient flight hours. We assume that in non-commercial airspaces, flight regulations would trust a certified pilot with sufficient reputation (i.e., no significant history of problems) to perform safety checks consistent with the lower-level details of our safety cases. In other words, the pilot is in charge of ensuring a safe flight in whatever airspace they are in. Regulators often do not exclude pilots legally allowed to be in the airspace unless there is some serious prior issue \cite{FAA_TRUST, FAA_part107}. So it is our belief that any UAS Traffic Management system will likely allow certified pilots to enter the airspace unless it has some reason not to.

As shown in Figure \ref{fig:pilot_only}, our safety case checks to see if the pilot is certified to fly their sUAS, here represented using the FAA's Part 107 certification \cite{FAA_part107}. We also check to see if the pilot has sufficient flight hours to be competent to complete this flight, which is something that our managed airspace should know.  In the future, this flight-hours check might be replaced or augmented with different checks, such as the pilot's score on a competency-reputation metric, future certifications, or temporary notices to pilots that the FAA might put out. If evidence of these checks confirms that the pilot is certified and has sufficient experience to enter the controlled airspace, the associated strategy node (S1) in the safety case is satisfied.

\begin{figure}[ht]
    \centering
    \includegraphics[width=\columnwidth]{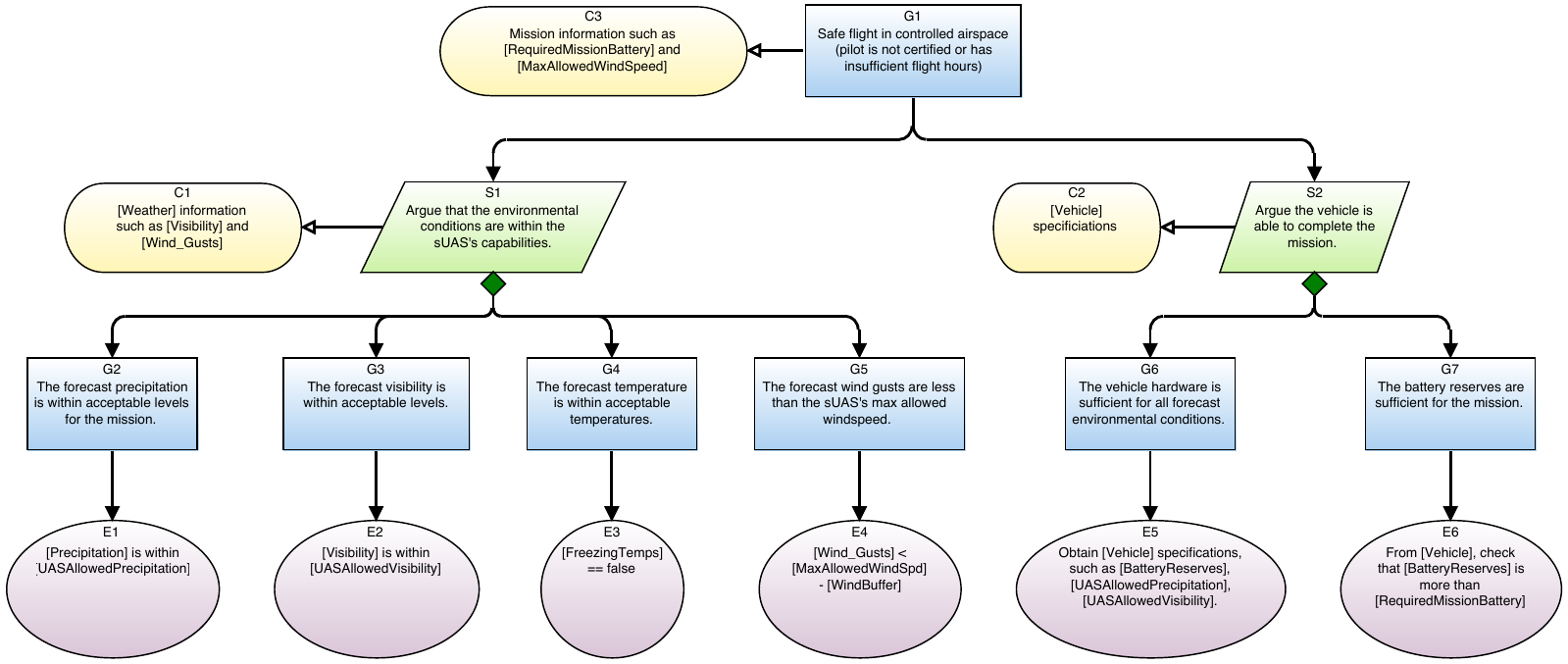}
    \caption{Wind Safety Case: A parameterized safety case based only on the weather and the drone's capabilities. This safety case creates the instances seen in Figures \ref{fig:instance_1} and \ref{fig:instance_2}.}
    \label{fig:wind_only}
\end{figure}

Our second safety case is relevant when the pilot lacks the evidence required to satisfy our initial safety case above. There needs to be an opportunity for newer pilots to learn and fly if such flights can be done safely. Thus, our second safety case focuses on giving such pilots the information that they will need in order to complete a safe flight. This second safety case (Figure \ref{fig:wind_only}) focuses on the weather because poor weather is a common reason for a pilot to decide that a flight will not be safe or for in-flight failures \cite{weather_hazards_for_UAV}. The weather portion of the feature model also has several parameters that can map to portions of our safety case. This sort of weather-focused safety case would normally involve far more attributes than we show in Figure \ref{fig:wind_only}, but we focus only on the weather and a small amount of information (evidence) about the battery here.

In the wind safety case from Figure \ref{fig:wind_only}, we constructed a general safety case that involves a number of parameters that are found in our feature model. These parameters are indicated using square brackets, such as [Precipitation] and [UASAllowedPrecipitation]. The data types of each parameter are left intentionally vague, as there are a number of ways for these parameters to be stored. We assume that information about each [Vehicle] is publicly available and that published sUAS specifications can be converted into the same data format and type that the feature model and safety case parameters have. If a [Vehicle] does not contain information in its specifications for certain parameters, then there is an option to assume some default values that could apply to almost all drones. 

For instance, most sUAS specifications will include information about the maximum allowed wind speed within which the manufacturer states the sUAS can operate. Likewise, most sUAS specifications include both maximum and minimum allowed temperatures in which to operate (often from -10 \textcelsius \;or 0 \textcelsius \;up to 40 \textcelsius) \cite{DEERCD20, DJI_MiniPro_4_Specs}. Fewer sUAS specifications contain specific information about visibility requirements since those depend on the type of mission being flown, especially whether it needs to be flown in a visual line of sight (VLOS) or beyond a visual line of sight (BVLOS). If the pilot does not provide visibility requirement information, we thus assume that the flight must take place VLOS and proceed accordingly. Similarly, if no information is provided about an sUAS's ability to fly in various forms of precipitation, we assume that the sUAS can only operate with no precipitation. 

Note that in the wind safety case (Figure \ref{fig:wind_only}), many of the goals share a similar structure. For instance, "The forecast precipitation is within acceptable level..." and "The forecast visibility is within acceptable levels...". The repetition of these elements is intentional and allows for greater ease of human understanding of the safety case, as well as for simpler extension of the safety case when we add additional hazards we need to mitigate. 

Some of the values of the parameters in the safety case may not be available at the time of a flight request. For instance, if a pilot is applying to complete a flight several weeks or months in the future, the forecast weather conditions will be unreliable. In such a case, the safety case might not contain concrete values until closer to the flight. The pilot could still access the parameterized safety case in order to study the safety requirements for the flight. As the time of the flight approaches, a more fully instantiated safety case could be sent to the pilot. 

The information for instantiating these parameterized safety cases will need to be pulled from a variety of sources, such as publicly available weather data and manufacturers' specifications for commercially available sUAS. However, some of the parameters' information will need to come from the pilot, including their certification status, the sUAS model they will fly, their flight plan, and any additional sUAS capabilities they have added (such as detect-and-avoid systems). 
In the event that the sUAS being flown was completely home-built, there may be no public documentation of its abilities, and all of its specifications will need to be provided (or inferred) by the pilot. 
Therefore, some of the individual safety cases will necessarily contain a fair amount of uncertainty while still serving as a guideline for the pilot.

\subsection{Instances}
\begin{figure}[ht]
    \centering
    \includegraphics[width=.95\textwidth]{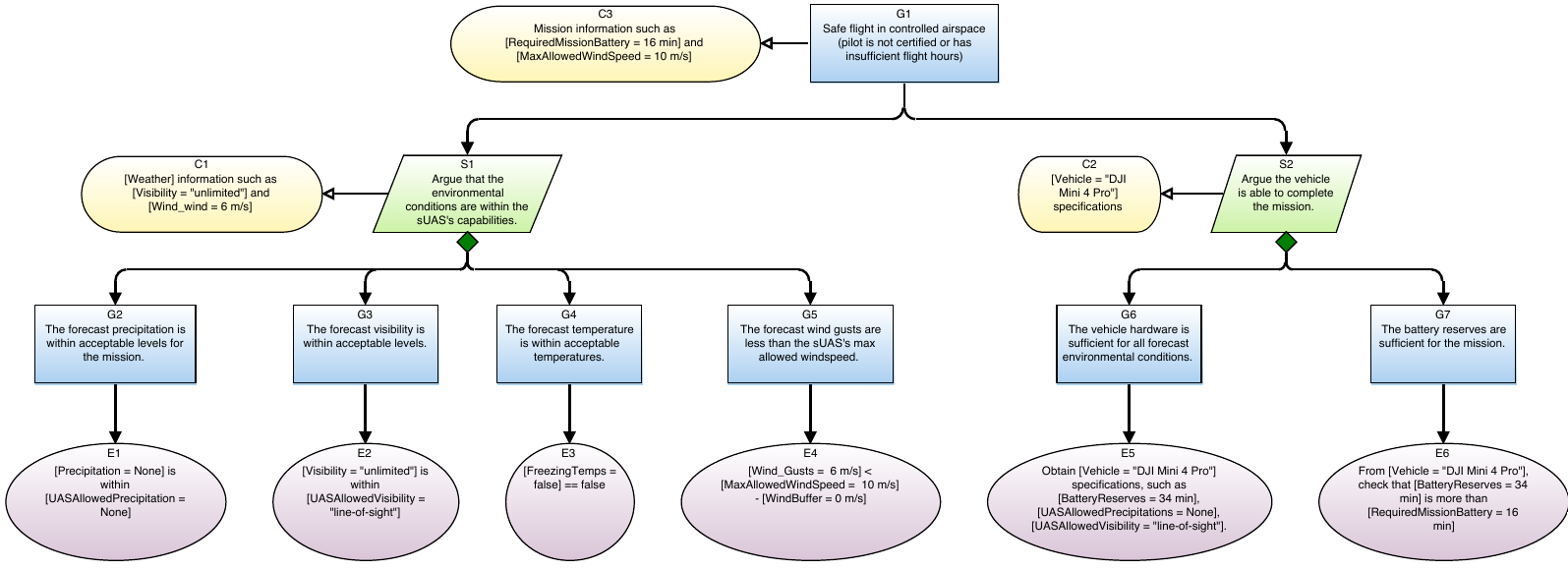}
    \caption{Safety Case Instance 1: An instance of the wind safety case (Figure \ref{fig:wind_only}) based on a mission with a DJI Mini 4 Pro drone.}
    \label{fig:instance_1}
\end{figure}

\begin{figure}[ht]
    \centering
    \includegraphics[width=.95\textwidth]{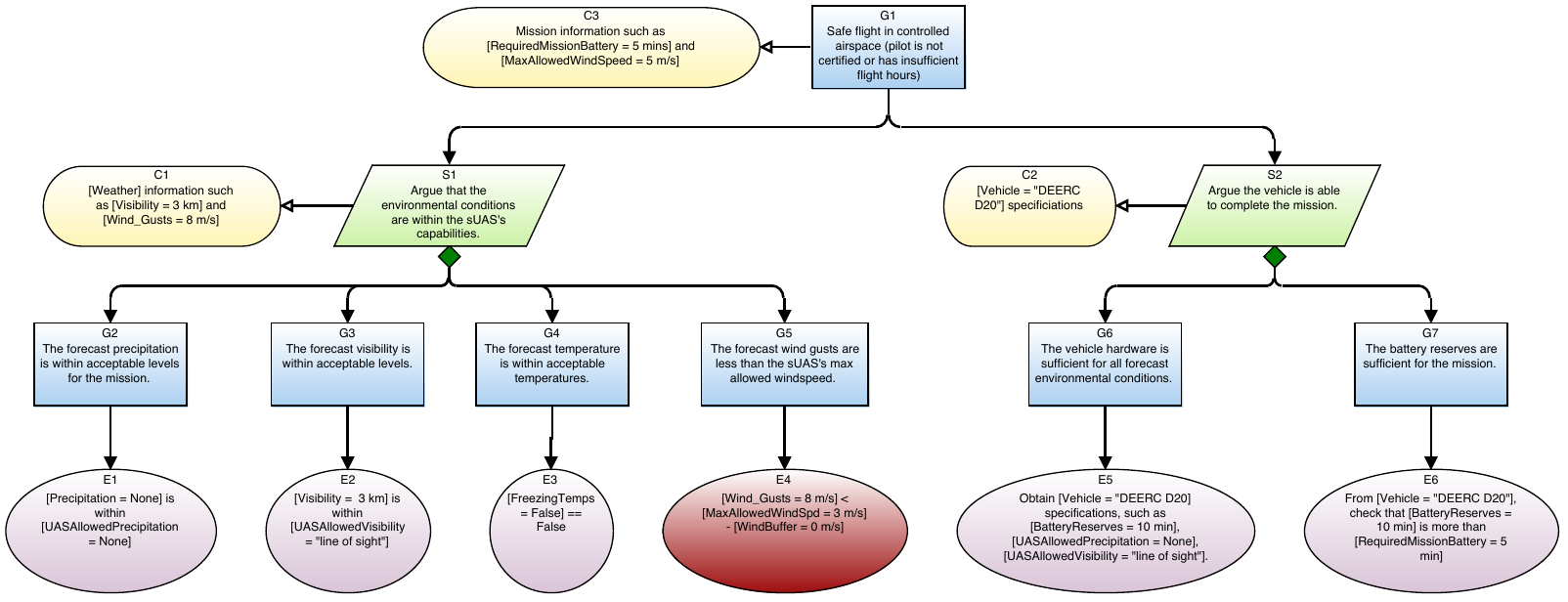}
    \caption{Safety Case Instance 2: An instance of the wind safety case (Figure \ref{fig:wind_only}) based on a mission with a DEERC D20 drone. Note that this instance fails to fulfill our safety requirements at node E4 (marked in darker red).}
    \label{fig:instance_2}
\end{figure}

As a final step in our SafeSPLE process, we demonstrate how to create instances of our parameterized safety case. This process involves obtaining the information required for all parameters and checking if all the solution nodes of the safety case remain true. In all of the safety case diagrams in Figures \ref{fig:pilot_only}, \ref{fig:wind_only}, \ref{fig:instance_1}, \ref{fig:instance_2}, these solution nodes are the bottom nodes labeled E1-E6, and have propagated from the context. If any solution node becomes false, then we can say that the pilot should either reconsider the flight, or should implement further mitigations to reduce the risk from the relevant hazard. For instance, if the safety case shows that the current wind gusts are too high, the pilot might delay the flight until the wind calms, or the pilot might decide to make the flight with a larger and more capable UAS (if available).

The first instance of our parameterized safety case is shown in Figure \ref{fig:instance_1}. This mission will be performed by a DJI Mini 4 Pro, a widely available drone that currently sells for just over \$1000, depending on accessories. The Mini 4 Pro is fully charged, and the mission, as planned, should take 16 minutes, flown entirely within VLOS of the pilot. This information about battery charge and the mission plan is provided by the pilot. The wind is gusting up to 6 meters/sec, with temperatures in the mid 20s \textcelsius, unlimited visibility, and no precipitation. This weather information is provided to the safety case by a commercial or governmental weather service. 

Once we obtain the information about the make and model of the drone, we can look up the DJI's published specifications. According to DJI \cite{DJI_MiniPro_4_Specs}, the Mini 4 Pro is able to fly in wind speeds up to 10 m/s, and with a fully charged battery can fly up to 34 minutes. The Mini 4 Pro can operate in temperatures between -10 \textcelsius \;and 40 \textcelsius. Using all this information, we can instantiate the safety case seen in Figure \ref{fig:instance_1}. Note that every solution node 
(labeled E1-E6) is satisfied by the above information. There is no precipitation; visibility is unlimited; the temperatures are not too hot or cold; the wind gusts are below the max allowed for the drone; and the battery reserves are more than twice as much as needed. So in this instance of the safety case the the top-level goal is satisfied. 

In Figure \ref{fig:instance_2} we can see a second instance of our safety case. This mission will be performed by a DEERC D20 drone, another widely available drone that currently sells for around \$50. The D20 is also fully charged, and the planned mission will only take 5 minutes of flying, entirely within VLOS. The wind is gusting up to 8 m/s, with temperatures in the mid-30s \textcelsius, 3 km visibility, and no precipitation.

According to the DEERC documentation \cite{DEERCD20}, the D20 drone is capable of about 10 minutes of flight time in temperatures between 0 \textcelsius \ and 40 \textcelsius. However, the D20 documentation does not specify the maximum speed of the winds that the drone is capable of flying in. Instead, the documentation reads, "DO NOT use this drone in adverse weather conditions such as rain, snow, fog, and wind." Therefore the safety case takes a conservative approach and assigns a default value of 3 m/s to the variable [MaxAllowedWindSpd] (3 m/s is slightly less than 7 mph). This default value could, of course, be set to 0 m/s, although this seems unrealistic for most outdoor flying. Other default values might be justified.

Plugging in all of these values, we see that while most solution nodes are satisfied, the current wind conditions (gusts up to 8 m/s) do not allow for a safe flight with the D20 (default max wind speed of 3 m/s). In Figure \ref{fig:instance_2}, this is shown at solution node E4, which is colored a darker red than the other solution nodes. The safety case is designed to serve as input to the UTM on-entry decision. At this point there are two main options for how the UAS Traffic Manager could behave. The UTM could refuse entry to this pilot until the wind speed is lower, or the UTM could send the safety case to the pilot with the recommendation that the pilot make modifications to the flight plan while leaving the ultimate flight decision up to the pilot.

Creating instances of safety cases with SafeSPL should be quick and relatively straightforward, if the information it needs is available. If information on the drone's capabilities is lacking, default values can still allow the safety case to create a reasonable instance. If information about the weather is unknown, then those portions of the safety case can be left uninstantiated until more detailed information becomes available. At the very least, we can generate a partially instantiated safety case so the pilot can see the areas where information is lacking or is based on default values. This information could allow the pilot to focus on mitigation measures in those areas if needed. 

\subsection{Connecting to Safe Entry}
\label{sec:safe_entry}

The parameterized safety cases created by SafeSPLE and described above could play an important role in a to-be-developed UTM system. When a pilot requests permission to fly in the airspace controlled by the UTM, the information needed to instantiate the safety case is either submitted by the pilot or looked up by the UTM system. Once a safety case has been created for that flight, there are at least two options for what the UTM system might do with it. 

\begin{enumerate}
    \item Closed Access: The UTM system accepts or denies requests based on whether each generated safety case "passes" or "fails". In other words, if the safety case goals are not satisfied, the UTM system  denies the flight. 
    \item Open Access: The UTM system accepts or denies the flight based solely on whether the pilot is certified or trusted. The safety case then becomes a guideline that can be provided to the pilot as something of a checklist to encourage a safer flight.
\end{enumerate}

Which action the UTM should take is an ongoing discussion with no immediate correct answer.
Currently the regulations in the US appear to generally favor approach (2), the open-access model. Regardless of which approach is taken for a specific controlled airspace, we believe the use of SafeSPLE will generate valuable on-the-fly information.  This information may offer an effective and useful checklist for decision-making. 

\section{Conclusion}

In this paper we have proposed the idea of SafeSPLE (or Safe Software Product Line Engineering) to support automation of related safety cases in controlled airspaces. We use a product line of features representing the expected context for a safety case. We then use parameterized safety cases where SPL features are mapped to the context and evidence nodes to  represent a family of (rather than individual) safety cases. Last, we will use the product line instances to instantiate individual safety cases on the fly, which can provide evidence for decision making regarding entry of an sUAS into a controlled airspace in an automated UTM. We have demonstrated our vision through a case study that focuses on two context nodes of a larger safety case--the pilot and weather. As future work we are building an automated tool for this process and plan to demonstrate SafePLE on a more complete safety case under a larger set of contexts.
\section{Acknowledgments}
The research described in this proposal is funded by NASA Grant Number: 80NSSC23M0058. We thank  Lilly Spirkovska for fruitful discussions and Ewen  Denney for providing the AdvoCATE tool.

\bibliography{main.bib}

\begin{thebibliography}{41}
\newcommand{\enquote}[1]{``#1''}
\providecommand{\natexlab}[1]{#1}
\providecommand{\url}[1]{\texttt{#1}}
\providecommand{\urlprefix}{URL }
\expandafter\ifx\csname urlstyle\endcsname\relax
  \providecommand{\doi}[1]{\discretionary{}{}{}https://doi.org/#1}\else
  \providecommand{\doi}[1]{\discretionary{}{}{}\urlstyle{rm}\url{https://doi.org/#1}}\fi

\bibitem[{Erdelj et~al.(2017)Erdelj, Natalizio, Chowdhury, and Akyildiz}]{Erdelj2017HelpFT}
Erdelj, M., Natalizio, E., Chowdhury, K.~R., and Akyildiz, I.~F., \enquote{Help from the Sky: Leveraging UAVs for Disaster Management,} \emph{IEEE Pervasive Computing}, Vol.~16, 2017, pp. 24--32.
\newblock \urlprefix\url{https://api.semanticscholar.org/CorpusID:18047608}.

\bibitem[{Administration(2023)}]{FAA}
Administration, F.~A., \enquote{UAS Sightings Report,} , June 2023.
\newblock \urlprefix\url{https://www.faa.gov/uas/resources/public_records/uas_sightings_report/}.

\bibitem[{Cleland-Huang et~al.(2022)Cleland-Huang, Chawla, Cohen, Al~Islam, Sinha, Spirkovska, Ma, Purandare, and Chowdhury}]{cleland2022towards}
Cleland-Huang, J., Chawla, N., Cohen, M., Al~Islam, M.~N., Sinha, U., Spirkovska, L., Ma, Y., Purandare, S., and Chowdhury, M.~T., \enquote{Towards Real-Time Safety Analysis of Small Unmanned Aerial Systems in the National Airspace,} \emph{AIAA AVIATION 2022 Forum}, 2022, p. 3540.
\newblock \doi{10.2514/6.2022-3540}.

\bibitem[{Michael and Gettinger(2017)}]{report3}
Michael, A.~H., and Gettinger, D., \enquote{Drone Incidents: A Survey of Legal Cases,} , Apr 2017.
\newblock \urlprefix\url{https://dronecenter.bard.edu/files/2017/04/CSD-Drone-Incidents.pdf}.

\bibitem[{Gohar et~al.(2024)Gohar, Hunter, Marczak-Czajka, Lutz, Cohen, and Cleland-Huang}]{gohar2024towards}
Gohar, U., Hunter, M.~C., Marczak-Czajka, A., Lutz, R.~R., Cohen, M.~B., and Cleland-Huang, J., \enquote{Towards Engineering Fair and Equitable Software Systems for Managing Low-Altitude Airspace Authorizations,} \emph{International Conference on Software Engineering: Software Engineering in Society ICSE-SEIS}, 2024, p. 177–188.
\newblock \doi{10.1145/3639475.3640103}.

\bibitem[{Denney and Whiteside(2012)}]{NASA12}
Denney, E., and Whiteside, I., \enquote{Hierarchical Safety Cases,} , Dec. 2012.
\newblock \urlprefix\url{https://ntrs.nasa.gov/api/citations/20130001737/downloads/20130001737.pdf}.

\bibitem[{Knight(2012)}]{Knig12}
Knight, J., \emph{Fundamentals of Dependable Computing for Software Engineers}, CRC Press, 2012.

\bibitem[{Hatcliff et~al.(2014)Hatcliff, Wassyng, Kelly, Comar, and Jones}]{Hatcliff14}
Hatcliff, J., Wassyng, A., Kelly, T., Comar, C., and Jones, P.~L., \enquote{Certifiably safe software-dependent systems: challenges and directions,} \emph{Proceedings of the on Future of Software Engineering, {FOSE} 2014, Hyderabad, India, May 31 - June 7, 2014}, 2014, pp. 182--200.
\newblock \doi{10.1145/2593882.2593895}.

\bibitem[{Leveson(2023)}]{Leveson23}
Leveson, N., \emph{An Introduction to System Safety Engineering}, MIT Press, 2023.

\bibitem[{Rakotonarivo et~al.(2023)Rakotonarivo, Drougard, Conversy, and Garcia}]{10.1145/3544548.3581003}
Rakotonarivo, B.~H., Drougard, N., Conversy, S., and Garcia, J., \enquote{Cleared for Safe Take-off? Improving the Usability of Mission Preparation to Mitigate the Safety Risks of Drone Operations,} \emph{Proceedings of the 2023 CHI Conference on Human Factors in Computing Systems}, 2023, pp. 1--17.
\newblock \doi{10.1145/3544548.3581003}.

\bibitem[{Prevot et~al.(2016)Prevot, Rios, Kopardekar, Robinson~III, Johnson, and Jung}]{UTM}
Prevot, T., Rios, J., Kopardekar, P., Robinson~III, J., Johnson, M., and Jung, J., \enquote{UAS Traffic Management (UTM) Concept of Operations to Safely Enable Low Altitude Flight Operations,} 2016.
\newblock \doi{10.2514/6.2016-3292}.

\bibitem[{Clements and Northrop(2001)}]{Clements2001}
Clements, P.~C., and Northrop, L., \emph{Software Product Lines: Practices and Patterns}, SEI Series in Software Engineering, Addison-Wesley, 2001.

\bibitem[{Weiss and Lai(1999)}]{weiss1999software}
Weiss, D., and Lai, C., \emph{Software product-line engineering: a family-based software development process}, Addison-Wesley, 1999.
\newblock \urlprefix\url{http://books.google.com/books?id=721YAAAAYAAJ}.

\bibitem[{Pohl et~al.(2005)Pohl, B\"{o}ckle, and van~der Linden}]{Pohl-PL-Eng-Book}
Pohl, K., B\"{o}ckle, G., and van~der Linden, F., \emph{Software Product Line Engineering: Foundations, Principles and Techniques}, Springer-Verlag New York, Inc., Secaucus, NJ, USA, 2005.

\bibitem[{Kang et~al.(1990)Kang, Cohen, Hess, Novak, and Peterson}]{Kang1990}
Kang, K.~C., Cohen, S.~G., Hess, J.~A., Novak, W.~E., and Peterson, A.~S., \enquote{Feature-Oriented Domain Analysis {(FODA)} Feasibility Study,} Tech. rep., Carnegie-Mellon University Software Engineering Institute, November 1990.

\bibitem[{{SEI, Software Engineering Institute}(2020)}]{SEI-PL}
{SEI, Software Engineering Institute}, \enquote{Software Product Lines,} \url{http://www.sei.cmu.edu/productlines}, 2020.

\bibitem[{Batory et~al.(2006)Batory, Benavides, and Ruiz-Cortes}]{DonBatory-FeatureDef}
Batory, D., Benavides, D., and Ruiz-Cortes, A., \enquote{Automated analysis of feature models: challenges ahead,} \emph{Commun. ACM}, Vol.~49, No.~12, 2006, pp. 45--47.
\newblock \doi{10.1145/1183236.1183264}.

\bibitem[{Kang et~al.(2002)Kang, Lee, and Donohoe}]{pl-feature-modeling-kang}
Kang, K.~C., Lee, J., and Donohoe, P., \enquote{Feature-Oriented Product Line Engineering,} \emph{IEEE Software}, Vol.~19, No.~4, 2002, pp. 58--65.
\newblock \doi{http://doi.ieeecomputersociety.org/10.1109/MS.2002.1020288}.

\bibitem[{Jacobson et~al.(1997)Jacobson, Griss, and Jonsson}]{DBLP:books/daglib/0087788}
Jacobson, I., Griss, M.~L., and Jonsson, P., \emph{Software reuse - architecture, process and organization for business}, Addison-Wesley-Longman, 1997.

\bibitem[{Gomaa(2004)}]{Gomaa}
Gomaa, H., \emph{Designing Software Product Lines with UML: From Use Cases to Pattern-Based Software Architectures}, Addison Wesley Longman Publishing Co., Inc., Redwood City, CA, USA, 2004.

\bibitem[{Fantechi and Gnesi(2007)}]{Fantechi-Gnesi}
Fantechi, A., and Gnesi, S., \enquote{A behavioural model for product families,} \emph{ESEC-FSE '07: Proc. of the Joint Meeting of the European Software Engineering Conf. and the ACM SIGSOFT Symp. on the Foundations of Software Engineering}, 2007, pp. 521--524.
\newblock \doi{10.1145/1287624.1287700}.

\bibitem[{Benavides et~al.(2010)Benavides, Segura, and Ruiz-Cortés}]{BENAVIDES2010615}
Benavides, D., Segura, S., and Ruiz-Cortés, A., \enquote{Automated analysis of feature models 20 years later: A literature review,} \emph{Information Systems}, Vol.~35, No.~6, 2010, pp. 615--636.
\newblock \doi{10.1016/j.is.2010.01.001}.

\bibitem[{Roos-Frantz et~al.(2012)Roos-Frantz, Galindo, Benavides, and Ruiz-Cort\'{e}s}]{fama}
Roos-Frantz, F., Galindo, J.~A., Benavides, D., and Ruiz-Cort\'{e}s, A., \enquote{{FaMa-OVM}: a tool for the automated analysis of OVMs,} \emph{Proceedings of the 16th International Software Product Line Conference - Volume 2}, 2012, p. 250–254.
\newblock \doi{10.1145/2364412.2364456}.

\bibitem[{Thüm et~al.(2014)Thüm, Kästner, Benduhn, Meinicke, Saake, and Leich}]{featureide}
Thüm, T., Kästner, C., Benduhn, F., Meinicke, J., Saake, G., and Leich, T., \enquote{FeatureIDE: An extensible framework for feature-oriented software development,} \emph{Science of Computer Programming}, Vol.~79, 2014, pp. 70--85.
\newblock \doi{10.1016/j.scico.2012.06.002}, experimental Software and Toolkits (EST 4): A special issue of the Workshop on Academic Software Development Tools and Techniques (WASDeTT-3 2010).

\bibitem[{Xiang et~al.(2018)Xiang, Zhou, Zheng, and Li}]{10.1145/3176644}
Xiang, Y., Zhou, Y., Zheng, Z., and Li, M., \enquote{Configuring Software Product Lines by Combining Many-Objective Optimization and SAT Solvers,} \emph{ACM Transactions on Software Engineering Methodology}, Vol.~26, No.~4, 2018.
\newblock \doi{10.1145/3176644}.

\bibitem[{Uzuncaova et~al.(2010)Uzuncaova, Khurshid, and Batory}]{5456077}
Uzuncaova, E., Khurshid, S., and Batory, D., \enquote{Incremental Test Generation for Software Product Lines,} \emph{IEEE Transactions on Software Engineering}, Vol.~36, No.~3, 2010, pp. 309--322.
\newblock \doi{10.1109/TSE.2010.30}.

\bibitem[{Firestone and Cohen(2018)}]{FirestoneC18}
Firestone, J., and Cohen, M.~B., \enquote{The Assurance Recipe: Facilitating Assurance Patterns,} \emph{Computer Safety, Reliability, and Security - {SAFECOMP} 2018 Workshops, ASSURE, DECSoS, SASSUR, STRIVE, and WAISE, V{\"{a}}ster{\aa}s, Sweden, September 18, 2018, Proceedings}, Lecture Notes in Computer Science, Vol. 11094, Springer, 2018, pp. 22--30.
\newblock \urlprefix\url{https://doi.org/10.1007/978-3-319-99229-7\_3}.

\bibitem[{Denney and Pai(2016{\natexlab{a}})}]{depai2016}
Denney, E., and Pai, G., \enquote{Composition of safety argument patterns,} \emph{Proc. of the Int'l Conf. on Computer Safety, Reliability, and Security}, Springer, 2016{\natexlab{a}}, pp. 51--63.
\newblock \urlprefix\url{https://doi.org/10.1007/978-3-319-45477-1\_5}.

\bibitem[{Vierhauser et~al.(2021)Vierhauser, Bayley, Wyngaard, Xiong, Cheng, Huseman, Lutz, and Cleland{-}Huang}]{VierhauserBWXCH21}
Vierhauser, M., Bayley, S., Wyngaard, J., Xiong, W., Cheng, J., Huseman, J., Lutz, R.~R., and Cleland{-}Huang, J., \enquote{Interlocking Safety Cases for Unmanned Autonomous Systems in Shared Airspaces,} \emph{{IEEE} Trans. Software Eng.}, Vol.~47, No.~5, 2021, pp. 899--918.
\newblock \doi{10.1109/TSE.2019.2907595}.

\bibitem[{Agrawal et~al.(2019)Agrawal, Khoshmanesh, Vierhauser, Rahimi, Cleland{-}Huang, and Lutz}]{AgrawalKVRCL19}
Agrawal, A., Khoshmanesh, S., Vierhauser, M., Rahimi, M., Cleland{-}Huang, J., and Lutz, R.~R., \enquote{Leveraging artifact trees to evolve and reuse safety cases,} \emph{Proceedings of the 41st International Conference on Software Engineering, {ICSE} 2019, Montreal, QC, Canada, May 25-31, 2019}, edited by J.~M. Atlee, T.~Bultan, and J.~Whittle, {IEEE} / {ACM}, 2019, pp. 1222--1233.
\newblock \doi{10.1109/ICSE.2019.00124}.

\bibitem[{{Assurance Case Working Group}(2021)}]{GSN}
{Assurance Case Working Group}, \enquote{Goal Structuring Notation Version 3,} , 2021.
\newblock \urlprefix\url{https://scsc.uk/SCSC-141C}.

\bibitem[{Denney and Pai(2018)}]{advocate}
Denney, E., and Pai, G., \enquote{Tool support for assurance case development,} \emph{Automated Software Engineering}, Vol.~25, No.~3, 2018, pp. 435--499.
\newblock \doi{10.1007/s10515-017-0230-5}.

\bibitem[{Ericson et~al.(2015)}]{ericson2015hazard}
Ericson, C.~A., et~al., \emph{Hazard analysis techniques for system safety}, John Wiley \& Sons, 2015.

\bibitem[{Denney and Pai(2016{\natexlab{b}})}]{denpai2016}
Denney, E., and Pai, G., \enquote{Architecting a Safety Case for UAS Flight Operations,} 2016{\natexlab{b}}.
\newblock \urlprefix\url{https://api.semanticscholar.org/CorpusID:19489937}.

\bibitem[{Clothier et~al.(2017)Clothier, Denney, and Pai}]{clodenpai2017}
Clothier, R., Denney, E., and Pai, G., \enquote{Making a Risk Informed Safety Case for Small Unmanned Aircraft System Operations,} 2017.
\newblock \doi{10.2514/6.2017-3275}.

\bibitem[{{EASA}(2024)}]{sora}
{EASA}, \enquote{Specific Operations Risk Assessment ({SORA}),} Available at \url{https://www.easa.europa.eu/en/domains/civil-drones-rpas/specific-category-civil-drones/specific-operations-risk-assessment-sora} (2024/06/23), 2024.

\bibitem[{FAA()}]{FAA_TRUST}
FAA, \enquote{The Recreational Uas Safety Test (trust),} , ????
\newblock \urlprefix\url{https://www.faa.gov/uas/recreational_flyers/knowledge_test_updates}.

\bibitem[{Administration(2024)}]{FAA_part107}
Administration, F.~A., \enquote{Become a Certificated Remote Pilot,} , February 2024.
\newblock \urlprefix\url{https://www.faa.gov/uas/commercial_operators/become_a_drone_pilot}.

\bibitem[{Averyanova and Znakovskaja(2021)}]{weather_hazards_for_UAV}
Averyanova, Y., and Znakovskaja, E., \enquote{Weather Hazards Analysis for small UASs Durability Enhancement,} \emph{2021 IEEE 6th International Conference on Actual Problems of Unmanned Aerial Vehicles Development (APUAVD)}, 2021, pp. 41--44.
\newblock \doi{10.1109/APUAVD53804.2021.9615440}.

\bibitem[{DEERC(2019)}]{DEERCD20}
DEERC, \enquote{DEERC D20 Instructions for Use v4.0,} , 2019.
\newblock \urlprefix\url{https://deerc.com/Uploads/Download/2022-11-11/636dac7c6f6d5.pdf}.

\bibitem[{DJI(2023)}]{DJI_MiniPro_4_Specs}
DJI, \enquote{DJI Mini 4 User Manual,} , 2023.
\newblock \urlprefix\url{https://dl.djicdn.com/downloads/DJI_Mini_4_Pro/DJI_Mini_4_Pro_User_Manual_EN.pdf}.

\end{thebibliography}

\end{document}